# Economic uncertainty and exchange rates linkage revisited: modelling tail dependence with high frequency data

**Nourhaine NEFZI[a]    Abir ABID[b,c]**

[a] MACMA, Institut Supérieur de Gestion de Tunis, Université de Tunis, Tunisie.
Email : Nourhaine.Nefzi@isg.rnu.tn

[b] LEO, Université d'Orléans (CNRS, FRE 2014), Rue de Blois, 45067 Orléans Cedex 2, France
[c] MACMA, Institut Supérieur de Gestion de Tunis, Université de Tunis, Tunisie.

Email : abir.abid@etu.univ-orleans.fr

**Abstract**

The aim of this paper is to dig deeper into understanding the exchange rates and uncertainty dependence. Using the novel Baker *et al.* (2020)'s daily Twitter Uncertainty Index and BRICS exchange rates, we investigate their extreme tail dependence within an original time-varying copula framework. Our analysis makes several noteworthy results. Evidence for Indian, Russian and South African currencies indicates an elliptical copulas' dominance implying neither asymmetric features nor extreme movements in their dependence structure with the global economic uncertainty. Importantly, Brazilian and Chinese currencies tail dependence is upward trending suggesting a safe-haven role in times of high global economic uncertainty including the recent COVID-19 pandemic. In such circumstances, these markets offer opportunities to significant gains through portfolio diversification.

**JEL** Classification: C22, E44, G15.

## 1. Introduction

The economic uncertainty (EU hereafter) is an amorphous phenomenon that reflects the situation in which agents are conscious of their little knowledge about whether economic conditions will change. Since the World economy has experienced a high level of EU throughout the recent years, numerous studies, who essentially build on the pioneer contribution of Bloom (2009), have empirically assessed its implications on all economic fields. In fact, there is a broad consensus on harmful economic impacts of uncertainty such as exacerbating financial frictions and lowering aggregate demand (Karaman and Yildirim-Karaman, 2019). A notable branch of this literature, that has particularly concerned itself with the asset pricing implications of uncertainty, establishes that an elevated EU lowers asset prices and rises market volatility (Pástor and Veronesi, 2012; Brogaard and Detzel, 2015)[1]. Though the stock market is at the core of this debate, the foreign exchange market has not garnered the same deal of attention. Interestingly, some papers argue that a rise in the economic policy uncertainty (EPU hereafter) (i) implies an increase of the exchange rate forecast errors (Beckmann and Czudaj, 2017), (ii) depreciates the domestic currency against the US dollar (Kido, 2016; Abid, 2019, Nilavongse *et al.* 2020) and (iii) imparts higher currency volatility (Krol, 2014; Bartsch, 2019; Zhou *et al.* 2019) for emerging as well as developed markets[2]. The literature also provides evidencing support to a negative uncertainty spillover from one country on exchange rates movements abroad (*Cf.* Nilavonsge *et al.* 2020; Bratsch, 2019)[3].

In spite of the confirmed empirical linkage, the extent to which currencies movements are associated with uncertainty is, however, not clear enough. This lack of clarity belongs essentially to whether their extreme co-movements are significant. Specifically, since exchange rate returns are known to usually exhibit asymmetric movements and fat-tails, looking only at their absolute dependence with the EU probably provides poor insight regarding their relationship. Then, investigating the dependence structure might help describe the ways in which the two variables are associated, especially because traditional correlation measures only quantify the strength of the relationship. Against the limited literature on this issue, our main purpose is to revisit the exchange rate-EU nexus by originally modelling their tail dependence structure in context of BRICS (Brazil, Russia, India, China and South Africa) countries. At a general level, the choice of emerging market economies (EME hereafter) is based around their vulnerability to external shocks including foreign uncertainty (*e.g.* Fernandez-Villaverde *et al.* 2011)[4]. Moreover, the BRICS block is particularly quite representative of the EME and well positioned as a powerful economic force, given its greater contribution to the global output as well as to the world trade[5]. Therefore, our study contributes to the recent debate in several fronts.

Our first contribution relates to analyzing the dependence structure that goes beyond linear association, an issue not yet explored in the applied literature. As an innovation, we implement, towards this end, a time-varying copula approach used for the first time in this context. To our best knowledge, the only related empirical work is Al-Yahyaee *et al.* (2020) who rely on quantile approach to investigate the exchange rate and EPU dependence. Despite separating recession and expansion phases, and upper and lower quantiles, Al-Yahyaee *et al.* (2020) have omitted the dependence structure as captured by copula, along with time-varying evolution of the considered pairs. So, what makes copula modelling relevant for our study? As argued by the literature, this method remains more flexible than other related tools to capture a variety of dependence structures which are particularly helpful for market participants. To ensure a relevant adjustment of their portfolio allocation and hedging decisions, they especially need to clearly identify periods when currencies are highly correlated to uncertainty which has not been explicitly identified in Al-Yahyaee *et al.* (2020). Unlike this study, our EGARCH –copula framework appears more appropriate, on the one hand, in encircling the intrinsic stylized facts of time series and on the other hand, in properly modelling the average and extreme interdependence. Moreover, particularly motivated by the renewed interest in identifying new safe-haven assets, we interestingly explore whether this role could be played by BRICS currencies in high global uncertainty episodes[6]. As such, market participants and policy authorities would extend their limited understanding of the ability of emerging currencies to act as a safe-haven against EU.

---

[1] Note that a key role is attributable to the uncertainty on government policy making process.
[2] The EPU refers to the situation in which economic agents are not able to clearly and fully foresee the future evolution of government economic policy actions (Baker *et al.* 2016).
[3] A special focus goes on the US EPU spillover.
[4] We emphasize that most of the papers dealing with uncertainty impacts on exchange rates are mainly applied to developed markets or Chinese data. To date, The BRICS group has not been the subject of a related investigation.
[5] See for example Gupta and Sun (2020) regarding the BRIC countries.

In additionally, our second contribution belongs to the frequency of the data used. To date, and excepting the notable research of Bartsch (2019), most of the empirical studies on currencies and uncertainty relationship are performed on a monthly basis[7]. Assuming that the data generating process of both uncertainty and exchange rates happens at a higher frequency, we embed the path of Bartsch (2019) and interestingly fill the gap of the literature by considering daily spot rates in response to daily economic uncertainty[8]. If we are not mistaken, our paper is again the first to exploit daily data in EME context.

Finally, our third contribution pertains to the measuring of the EU. Though increased attention goes on EPU implications on currencies movements, we take a step further from these works by considering uncertainty covering different economic areas. This is based on the conjecture that the EPU fails to describe the whole picture of the EU. To capture the global uncertainty, we rely on the recently developed Twitter Economic Uncertainty (TEU, in short) of Baker *et al.* (2020) that appears reasonably well positioned to reflect the global EU. Specifically, this measure is a daily coverage of twitter messages including terms related to fields such as "economics", "finance", "equity", "trade", "investments"[9]. Because the speed at which information spreads on social media exceeds the written press, the TEU index is likely to provide information beyond the extensively used news-papers based uncertainty index of Baker *et al.* (2016)[10]. To our best knowledge, the TEU index is used for the first time to examine the asset pricing implications of uncertainty.

Our empirical evidence reveals various dependence structures between the global EU and the BRICS currencies. Through scrutinizing the tail dependence behavior, we particularly identify upward movements of the Chinese and Brazilian currencies along elevated EU episodes. Relatedly, we conclude a safe-haven role of these currencies against uncertainty that implies interesting diversification opportunities.

We describe the data and econometric design in Section 2, followed by the empirical findings in Section 3. Section 4 briefly discusses a robustness analysis and Section 5 concludes the work.

## 2. Data and Econometric strategy

The dataset in question is composed of the daily spot exchange rates of the BRICS national currencies per the US dollar and the daily TEU discussed above[11]. The former are retrieved from DataStream while the latter is sourced from www.policyuncertainty.com. We start our sample in January 2011, date at which the uncertainty index is made available, while the ending date is in all cases June 30, 2020. Beyond the data difficulties, the examined period interestingly covers major events that trigger higher uncertainty: the euro debt crisis (summer, 2011); the loss of the AAA notation in the USA (2012), the Chinese stock market Karsh (2015), the announcement of the Brexit referendum (2016), the president Trump's election (2016), the Sino-American trade conflict, and specifically the COVID-19 pandemic (January, 2020 to present)[12].

---

[6] Note that in the later part of the paper, the economic merits of copula method are also assessed through investigating the dynamic association between the EU and the Japanese currency. Doing so, we test the ability of the copula framework to highlight the well-known safe-haven nature of the Japanese Yen in times of higher US uncertainty.
[7] Bartsch (2019) argues that low frequency exchange rate data poorly detects the uncertainty impacts.
[8] Probably, the major drawback of Bartsch (2019) is the examining the US dollar per British pound exchange rate alone.
[9] The Baker *et al.* (2020) TEU covers all twitter messages in English language (both inside and outside the USA). However, the authors argue that the TEU is by far a measure of the US economic uncertainty since more than two-thirds of twitter users of all English speaking countries are located in the USA. As the US uncertainty is widely perceived as a benchmark for the global uncertainty, we consider the TEU index as an appropriate proxy for the global EU.
[10] Note that we have examined the US EPU and the TEU correlation. The result reveals a 0.5 Pearson coefficient. This gives support to our initial intuition as the EU is reasonably explained by other sources of uncertainty beyond the EPU.
[11] BRL/USD; RUB/USD; INR/USD, CHY/USD and ZAR/USD are retained respectively for the Brazilian Real, Russian Ruble, Indian Rupee; Chinese Yuan, and South African Rand against the US Dollar.
[12] Ahir *et al.* (2018) postulate that a large part of the world uncertainty is attributable to developed countries uncertainty. The role of the Brexit in increasing the TEU index is clear as can be seen in the Figure A.1 of the Appendix A. In addition, Baker *et al.* (2020) document a surge of global economic uncertainty throughout the period of the COVID-19 pandemic.

Because both exchange rates and uncertainty evolutions are known to be asymmetric, the assumption of multivariate normal distribution is likely to be not appropriate to describe the dependence structure between the two variables. With this in mind, we filter the time series using appropriate ARMA (m,n)– EGARCH(p,q) processes as to precisely describe their dynamic dependence structure[13]. In the next step, a number of copula functions are fitted to the filtered series in order to gauge their dynamic linkage. Further description of our methodology is presented in the following sub-sections.

**(i)   Marginal Distribution**

As previously mentioned, the first empirical exercise is to model the marginal distribution of each used time series that we denote $Z_t$. To this end, an ARMA-EGARCH model with GED distribution is considered.

$$Z_t = u_t + \varepsilon_t = u_t + \sqrt{h_t}\mu_t \tag{1}$$

, where $u_t = a_0 + \sum_{i=1}^{m} a_i Y_{t-i} + \sum_{j=1}^{n} b_j \varepsilon_{t-j}$; $a_0$ is a constant; $a_i$ and $b_j$ are respectively, the autoregressive (AR) and the moving average (MA) polynomials; $Z_{t-i}$ corresponds to either the lagged exchange return of each country or the EU index; $\varepsilon_t$ are the residuals; $\mu_t = \varepsilon_t / \sqrt{h_t}$ is a normalized residual ; and $h_t$ is a conditional variance of $\varepsilon_t$. In this vein, $h$ is supposed to be an EGARCH process (Nelson, 1991) described in the Eq. (2)[14]:

$$\log h_t = w + \sum_{i=1}^{p} \varkappa_i \frac{|\varepsilon_{t-i}| + \gamma_i \varepsilon_{t-i}}{\sqrt{h_{t-i}}} + \sum_{j=1}^{q} \Upsilon_j \log h_{t-j} \tag{2}$$

, where $w$ is a constant, and $\varkappa$ reflects the magnitude effect in the GARCH model (ARCH effect). GARCH and asymmetry effects are respectively captured by "$\Upsilon$" and "$\gamma$".

**(ii)   The copula models for the joint distributions**

According to the copula theorem (Sklar, 1959), any n-dimensional joint distribution function can be factored into marginal distribution and a dependence structure (C). For two random variables X and Y with respectively F and G continuous distributions, and H a joint distribution of the couple (X, Y), there is a single copula that could be modelled as follow:

$$C: H(X,Y) = C(u,v) \tag{3}$$

$C$ is used to capture the dependence structure between uniform variables $u = F(X)$ and $v = G(Y)$. The literature has emphasized the effectiveness and the flexibility of copula functions in providing valuable information on the average (absolute) and tail (extreme) dependences. This latter is measured by the upper and the lower tail dependence coefficients.

$$\lambda_U(\alpha) = \lim_{\alpha \to 1^-} P(Y > G^{-1}(\alpha) | X > F^{-1}(\alpha)) = \lim_{\alpha \to 1^-} \frac{1 - 2\alpha + C(\alpha,\alpha)}{1-\alpha}; \ \lambda_U \in [0,1) \tag{4}$$

$$\lambda_L(\alpha) = \lim_{\alpha \to 0^+} P(Y < G^{-1}(\alpha) | X < F^{-1}(\alpha)) = \lim_{\alpha \to 0^+} \frac{C(\alpha,\alpha)}{\alpha}; \ \lambda_L \in [0,1) \tag{5}$$

, where, $\lambda_U$ (upper tail dependence) and $\lambda_L$ (lower tail dependence) respectively quantify the probability that the two variables jointly experience extreme upwards and downwards. Using these coefficients, we can capture the tendency where extremely high (low) global EU occurs simultaneously with currency high (low) returns. However, as exchange rate returns are sensitive to global market and political conditions, considering constant parameters in the dependence distribution may distort the interdependence features between them. We consider therefore the extended model of Patton (2006) who introduces the conditional (dynamic) copula. Unlike static copulas, this methodology captures the time variation in the dependent structure, which makes sense given that variables considered in this study are widely known to change over time.

---

[13] By applying this model, we will be able to deal with the autocorrelation and heteroscedasticity problems that our data contain as well as some well-known empirical stylized facts in asset returns, such as, the fat tail, asymmetric movement of volatility and leverage effect (see Table 1).

[14] We use the log return of all considered exchange rates.

### (iii) Dynamic Copula Functions

Several forms of copulas either Elliptical or Archimedean could be applied in finance[15]. In our study, we rely on five bivariate copula functions that are commonly used in the literature.

**a- Normal Copula** (has no tail dependence, $\lambda_U = \lambda_L = 0$):

$$(u,v|\rho) = \int_{-\infty}^{\Phi^{-1}(u)} \int_{-\infty}^{\Phi^{-1}(v)} \frac{1}{2\pi\sqrt{(1-\rho^2)}} \exp\left\{\frac{-(r^2 - 2\rho rs + s^2)}{2(1-\rho^2)}\right\} drds, \rho \in (-1, 1) \tag{6}$$

$$= \Phi_\rho(\Phi^{-1}(u), \Phi^{-1}(u)(v)) \tag{7}$$

, where $\rho$ is the dependence parameter (*i.e.* the linear Pearson's correlation coefficient). $\Phi^{-1}(.)$ denotes the inverse of the distribution function of the standard normal. $\Phi_\rho$ is the bivariate cumulative distribution of the standard normal. The dynamic evolution of the normal copula parameters (Patton, 2006), is expressed as follows:

$$\rho_t = \Lambda\left(\omega_N + \beta_N \rho_{t-1} + \alpha_N \cdot \frac{1}{10}\sum_{j=1}^{10} \Phi^{-1}(u_{t-j}) \cdot \Phi^{-1}(v_{t-j})\right) \tag{8}$$

, where $\Lambda$ is a logistic transformation that aims to keep the dependence parameter in a fixed interval. $\omega(.)$ is a constant parameter. $\beta(.)$ is the autoregressive term and $\alpha(.)$ is a 'forcing' variable that captures the variation in the dependence structure and defined as the absolute difference between $u_t$ and $v_t$ for ten previous observations.

**a- Student Copula** (shows symmetric tail dependence)

$$C_T(u,v|\rho,\vartheta) = \int_{-\infty}^{t_\vartheta^{-1}(u)} \int_{-\infty}^{t_\vartheta^{-1}(v)} \frac{1}{2\pi\sqrt{(1-\rho^2)}} \exp\left\{1 + \frac{(r^2 - 2\rho rs + s^2)}{\vartheta(1-\rho^2)}\right\}^{-\frac{\vartheta+2}{2}} drds, \tag{9}$$

This copula is able to detect extreme return co-movements via the symmetric parameter $\rho$ and degrees of freedom parameter $\vartheta$. The time-varying structure of student-t parameters are as follows:

$$\rho_t = \Lambda\left(\omega_{1T} + \beta_{1T}\rho_{t-1} + \alpha_{1T} \cdot \frac{1}{10}\sum_{j=1}^{10} T_\vartheta^{-1}(u_{t-j}) \cdot T_\vartheta^{-1}(v_{t-j})\right) \tag{10}$$

$$\vartheta_t = \tilde{\Lambda}\left(\omega_{2T} + \beta_{2T}\vartheta_{t-1} + \alpha_{2T} \cdot \frac{1}{10}\sum_{j=1}^{10} T_\vartheta^{-1}(u_{t-j}) \cdot T_\vartheta^{-1}(v_{t-j})\right) \tag{11}$$

### *b- Gumbel Copula*

This function describes only the upper tail dependence, $\theta$ and thus has zero lower-tail dependence. Gumbel copula is therefore able modelling the joint increase of the two variables.

$$C_G(u,v|\theta) = exp\left(-(-\log u)^\theta + (-\log v)^{\theta^{1/\theta}}\right), \theta \in [1, \infty) \tag{12}$$

The dynamic structure of Gumbel parameter is as below:

$$\theta_t = \Lambda\left(\omega_G + \beta_G \theta_{t-1} + \alpha_G \cdot \frac{1}{10}\sum_{j=1}^{10}|u_{t-j} - v_{t-j}|\right) \tag{13}$$

### *c- Clayton Copula*

This copula only exhibits lower tail dependence $\delta$ while assuming no dependency in the upper tail, *i.e.,* describes the phenomenon of a simultaneous dramatic decrease of the two variables.

$$C_C(u,v|\delta) = (u^{-\delta} + v^{-\delta} - 1)^{-1/\delta}, \delta \in (0, \infty) \tag{14}$$

, where the parameter $\delta_t$ has the following dynamic structure

$$\delta_t = \Lambda\left(\omega_C + \beta_C \delta_{t-1} + \alpha_C \cdot \frac{1}{10}\sum_{j=1}^{10}|u_{t-j} - v_{t-j}|\right) \tag{15}$$

---

[15] Elliptical copulas have the ability to model symmetric dependency structure while Archimedean copulas allow capturing asymmetric tail dependence.

### d- Symmetrised Joe-Clayton (SJC) copula

This function is able to simultaneously model upper and lower tail dependence, $\lambda_U$ and $\lambda_L$ respectively, thus allowing accounting for asymmetric dependence.

$$C_{SJC}(u,v|\lambda_U,\lambda_L) = \frac{1}{2}(C_{JC}(u,v) + C_{JC}(1-u,1-v) + u + v - 1) \quad (16)$$

, where $C_{JC}$ is the Joe-Clayton copula having the following expression:

$$C_{SJC}(u,v|\lambda_U,\lambda_L) = 1 - \left(1 - \{[1-(1-u)^a]^{-b} + [1-(1-v)^a]^{-b} - 1\}^{-\frac{1}{b}}\right)^{\frac{1}{a}} \quad (17)$$

With $a = \frac{1}{\log_2(2-\lambda_U)}$; $b = -\frac{1}{\log_2(\lambda_L)}$ and $(\lambda_U, \lambda_L) \in (0,1)$

The time-varying structure of $\lambda_U$ and $\lambda_L$ which range freely and do not dependent on each other, could be modelled as follow[16]:

$$\lambda_U = \Lambda\left(\omega_U + \beta_U \lambda_{U_{t-1}} + \alpha_U \cdot \frac{1}{10}\sum_{j=1}^{10}|u_{t-j} - v_{t-j}|\right) \quad (18)$$

$$\lambda_L = \Lambda\left(\omega_L + \beta_L \lambda_{L_{t-1}} + \alpha_L \cdot \frac{1}{10}\sum_{j=1}^{10}|u_{t-j} - v_{t-j}|\right) \quad (19)$$

Because each of the copulas has specific characteristics of tail dependence, we need to specify which copula provides the best representation of the tail dependence structure between BRICS currencies and the global EU. To achieve this goal, the related literature conventionally uses likelihood ratio tests or information criteria such as Akaike (AIC) or Schwarz's Bayesian Information Criteria (BIC). In line with the literature, we consider AIC rather than likelihood ratio tests on consideration to its optimality properties and ability to compare nested to non-nested models.

### 3. Empirical findings and Discussion

We start by the analysis of the summary statistics (not reported here in order to save space) of the TEU index and the log currency returns. We find (i) negative average returns of all currencies, (ii) an unconditional volatility led by the Russian currency followed by the Brazilian currency, (iii) positive Skewness for the TEU indicating probable uncertainty increase and (iv) highly kurtosis (exceeding 3) suggesting the presence of extreme values in the data. In addition, the Jarque-Bera test fully confirms the stylized non normality of the data. The Ljung-Box and Engle's ARCH LM tests respectively highlight serial dependence and heteroscedasticity. Given that, GARCH models are well suited tools to better capture these well-known stylized features.[17] Finally, Pearson's correlation indicates negative association between currency returns and the EU. However, because of its linear assumption, it is neither appropriate for providing insight into currency performance during high/low EU nor robust for assessing nonlinear behavior.

Moving now to the estimated results of ARMA (m,n) EGARCH(p,q), the following specifications are retained for the exchange rate return: EGARCH (1,2) (BRL/USD and ZAR/USD); EGARCH(4,2) (CNY/USD); AR(2)-EGARCH(1,3) (RUB/USD) and ARMA(1,1) EGARCH(2,1) (INR/USD). The EU is fitted via ARMA (2, 1) EGARCH (1, 1)[18].
Excepting the Chinese currency, there is strong evidence of GARCH effects and leverage effect indicating, respectively, the persistent of volatility over time for all considered variables as well as the higher impact of bad news on volatility relative to good news. LM, Q(30) and $Q^2$(30) tests confirm that standardized residual and squared standardized residuals don't suffer neither from heteroscedasticity nor from autocorrelation. These findings suggest that the estimated models are well specified enough to describe the patterns of the different variables and the extracted standardized residuals are independently and identically distributed.

---

[16] SJC copula becomes symmetric only if $\lambda_U = \lambda_L = 0$.
[17] The stationarity property is checked via Augmented Dickey-Fuller (ADF) and Philips-Perron (PP) tests. Both tests confirm the stationarity of all the variables. The results are not reported here but will be available upon request.
[18] To save space, the ARMA-EGARCH estimates are not reported here but are available from the authors upon request. We highlight that the retained ARMA specification (*i.e.* m and n parameters), is fixed based on the autocorrelation that we want to eliminate from the standardized and squared standardized residuals.

On the basis of filtered residuals, we estimate static and time-varying bivariate copulas for BRICS currencies and TEU index pairs. The corresponding results are reported in Table 1. Because the SJC copula parameters don't exhibit any statistical significance, they are not reported here. A first look at the results (AIC values) enables us to conclude that elliptical copulas have a better fit for RUB-EU; INR-EU and ZAR-EU pairs. More specifically, we find strong evidence of time-varying dependence of the Indian currency since, among all copulas; the Normal Copula displays the lowest AIC value. The RUB-EU fails to confirm a tail dependence structure. Moreover, we provide evidence on a symmetric tail dependence structure of the South African currency with the EU (based on the performance of the Student-t Copula in the static manner). Interestingly, evidences from Brazilian and Chinese cases strongly support Gumbel copula as the best fitting one. All estimated coefficients are statistically significant confirming the presence of upper tail dependence with the EU. From an economic point of view, this finding reveals that, when global EU reaches a high level, both currencies are likely to appreciate against the US dollar. This confirms that Chinese and Brazilian currencies act as a safe-haven asset in times of higher uncertainty. Our evidence provides interesting new insight alongside previous works since only the Japanese Yen has been empirically found as a safe-haven currency in times of higher US EPU (Kido, 2016; Beckmann and Czudaj, 2017).

Table 1. Static and Time-varying Copula Estimates of the Global Economic Uncertainty and BRICS currencies

| | BRL | RUB | INR | CNY | ZAR | BRL | RUB | INR | CNY | ZAR |
|---|---|---|---|---|---|---|---|---|---|---|
| **Panel A. Static copulas** | | | | | | | | | | |
| | | | Normal | | | | | Student-t | | |
| CST | -0.0028*** | -0.034*** | -0.03*** | -0.006*** | -0.026 | -0.00 | -0.031 | -0.03 | -0.006 | -0.025 |
| | [0.00045] | [0.00044] | [0.00044] | [0.00042] | [0.0004] | [0.02] | [0.0216] | [0.0216] | [0.021] | [0.022] |
| DF | – | – | – | – | – | 39.3 | 199.8*** | 199.9*** | 199.9*** | 24.5** |
| | | | | | | (---) | (9.78) | (8.4) | (3.95) | (12.3) |
| AIC | -0.27 | **-3.2** | -3.58 | -0.57 | -3.2 | 0.49 | -1.25 | -0.91 | 1.92 | **-4.8** |
| LL | 0.138 | 1.61 | 1.8 | 0.29 | 1.627 | -24.8 | 1.63 | 1.5 | 0.037 | 3.4 |
| | | | Gumbel | | | | | Clayton | | |
| CST | 1.004*** | 1*** | 1*** | 1*** | 1*** | -0.00 | -0.041** | -0.0378 | -0.008 | -0.024 |
| | [0.01] | [0.013] | [0.013] | [0.013] | [0.013] | [0.02] | [0.018] | [0.0257] | [0.026] | [0.0194] |
| AIC | 56.147 | 77.074 | 88.12 | 66.1 | 79.41 | 2.167 | 2.00 | 2.00 | 2.00 | 2.00 |
| LL | 28.07 | 38.53 | 44.06 | 33.1 | 39.7 | -0.13 | -0.00 | -0.00 | -0.00 | -0.00 |
| **Panel B. Time-varying copulas** | | | | | | | | | | |
| | | | Normal | | | | | Student-t | | |
| ω | – | – | – | – | – | 41.92 | 198.7 | 199.9 | 199.8 | 24.9 |
| | | | | | | (216.5) | (256.1) | (4858.4) | (138.8) | (159.5) |
| α | 0.0048*** | 0.0062*** | 0.023*** | 0.016*** | 0.00 | 0.005 | 0.006 | 0.02 | 0.016 | 0.00 |
| | [90.843] | [0.0090] | [0.011] | [0.01] | [7419.4] | [0.005] | [0.009] | [5.44] | [0.01] | [1.04] |
| β | 0.0004*** | 0.86*** | 0.84*** | 0.64*** | 0.28 | 0.97*** | 0.85*** | 0.84 | 0.64*** | 0.7 |
| | [----] | [0.05] | [0.07] | [0.1] | [---] | [0.036] | [0.07] | [31.4] | [0.14] | [2.5] |
| AIC | 3.64 | 0.31 | **-5.29** | 2.1 | 0.74 | 2.87 | 2.3 | -2.81 | 4.45 | -0.91 |
| LL | 0.18 | 1.84 | 4.65 | 0.95 | 1.63 | 1.56 | 1.85 | 4.405 | 0.774 | 3.46 |
| | | | Gumbel | | | | | Clayton | | |
| ω | -4.33*** | 3.17*** | 2.99*** | 3.38*** | 2.97*** | 0.00016 | 0.00014 | 0.000145 | 0.0002 | 0.00016 |
| | [0.143] | [0.071] | [0.005] | [0.024] | [0.33] | [2e-09] | [1e-08] | [3.7e-10] | [9e-09] | [1e-09] |
| α | 4.001*** | -3.14*** | -3.024*** | -3.265*** | -2.9*** | -1.65*** | -1.42*** | -1.44*** | -1.87*** | -1.61*** |
| | [0.13] | [0.062] | [0.004] | [0.02] | [0.28] | [6e-07] | [1e-06] | [7.1e-09] | [5e-07] | [3e-07] |
| β | 0.55*** | -0.052*** | 0.047*** | -0.33*** | -0.15 | -5e-06 | 2.7e-06 | -2.6e-06 | -1e-05 | -6e-06 |
| | [0.07] | [0.02] | [0.0015] | [0.0003] | [0.12] | [0.00] | [2e-09] | [2e-10] | [1e-09] | [3e-10] |
| AIC | **-1.74** | 2.3 | -2.04 | **-1.72** | 0.75 | 0.0086 | 0.022 | 0.024 | 0.024 | 0.014 |
| LL | -0.873 | 1.848 | -1.023 | -0.8621 | 1.627 | 0.0031 | 0.0097 | 0.011 | 0.011 | 0.006 |

Note: This Table illustrates the obtained results of static and time-varying copula model estimates. BRL, RUB, INR, CNY and ZAR denote the exchange rate of the national currency per the US dollar for Brazil, Russia, India, China and South Africa. CST={ρ,θ,δ} is the constant parameter of the copula. LL is log-likelihood. Values in bold reflect the lower AIC. ** and *** indicate statistical significances at 5% and 1% levels respectively. Standard errors are given in brackets. (---) denotes large number.

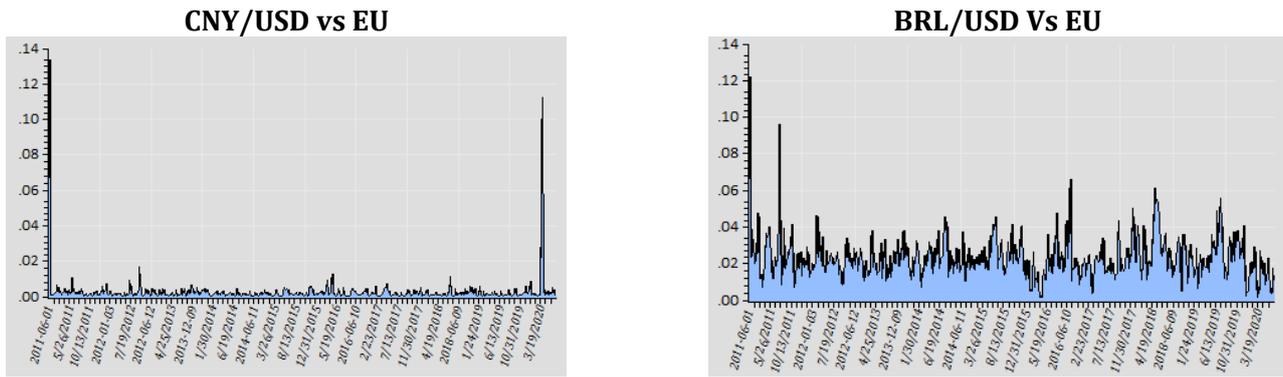

**Figure1. Exchange Rate Return Tail Dependence with the global EU (Gumbel Copula)**

The Figure.1 depicts the evolution path of the Gumbel tail dependence of EU with the BRL and the CNY. As can be clearly seen, both currencies show significant peaks throughout the whole sample period. The BRL and CNY appreciation is particularly pronounced in 2011-2012. A candidate explanation is that the period coincides with the debt crisis of the European countries which induces higher level of uncertainty and probably damages the attractiveness of European markets. We would expect that EME have benefited from these unusual economic conditions. Indeed, the equity inflows towards emerging markets increase in response to a surge of the European uncertainty. As a result, currency return improves. What is more, it is perhaps surprising that the CNY exhibits a significant appreciation along the first months of 2020 *i.e* the COVID-19 pandemic. Reasonably, this finding is somewhat counterintuitive as the origins of the pandemic are traced in China; however it is in line with Iqbal *et al.* (2020) showing a quite limited effects of COVID-19 shocks on Chinese currency. After all, China is one of the first to revive itself out of the crisis and has played a role of model to other countries.

4. **Sensitivity Analysis**

In order to increase the reliability of our findings, we analyze the dependence structure of the Japanese Yen (JBY/USD) and the global EU through copula functions. The choice of the Japanese currency is based on its confirmed prominent role as a safe-haven asset against the US uncertainty (Beckmann and Czudaj, 2017; Kido, 2016). We report the corresponding results in Table B.1 in Appendices. Interestingly, Gumbel copula is found to be the optimal fitting one to account for the dynamic dependence between global EU and JPY/USD exchange rate. Furthermore, positive tail dependence is clear from the results reported in Figure B.2 in Appendices. Accordingly, the Japanese Yen is likely to increase with higher level of global EU. Our result corroborates the previous literature and suggests that the copula methodology is well suited tool to account not only for the exchange rate and uncertainty non-linear dependence, but also for time variation in their dependence structure which can be economically advantageous.

5. **Conclusion**

In this paper, we investigated the patterns and trends in absolute and tail dependence between emerging markets currencies and global economic uncertainty (EU). Focusing on the BRICS case, we believe that our empirical strategy and concomitant conclusions provide novelties alongside the available literature. Our evidence showed various dependence structures: while elliptical copulas have been found as the best-fitting models for Russian; Indian and South African currencies, the Chinese Yuan and the Brazilian Real interactions with the global EU were better modelled by the Gumbel copula. The evidence we provide suggest that the BRICS currencies association to the global EU is not obvious during moderate uncertainty periods *i.e.* low uncertainty level (both SJC and Clayton copulas are found to be not significant). Based on this, we particularly corroborate the studies (*e.g.* Brograad and Detzel, 2015) suggesting more pronounced association of the asset prices to uncertainty during bad economic conditions (characterized by higher uncertainty).

Interestingly, our study provides the first comprehensive assessment of emerging currencies role as safe haven assets in times of higher uncertainty. In order to obtain significant gain through portfolio diversification, it would be relevant to investors to shift portfolio allocation into Brazilian and Chinese markets in periods of high global EU. Overall, we emphasizes that the EME are not as "less safe" as generally documented (*cf.* Bhattarai *et al.* 2019 for instance).

To complement our study, it would be important to delve deeper into understanding developed market currencies dependence to the global EU. Interesting econometric tools such as the copula framework would probably help seek for new safe-haven currencies.

# Appendices

## A. Economic Uncertainty Evolution

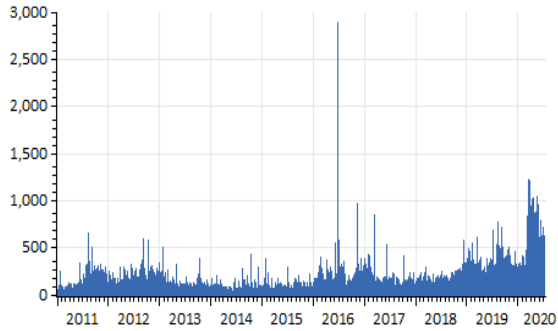

Figure A.1. Twitter Economic Uncertainty Index (01-01-2011/ 30-06-2020)

## B. Robustness Check

Table B.1. Copula Model Estimates of the Global Economic Uncertainty and Japanese currency

| | Static copula models | | | | | Dynamic copula models | | | |
|---|---|---|---|---|---|---|---|---|---|
| | Normal | t-student | Gumbel | Clayton | | Normal | t-student | Gumbel | Clayton |
| **CST** | 0.06 | 0.055** | 1.04*** | 0.005 | ω | - | 199.94 | -0.36*** | -3.46 |
| | [0.0201] | [0.021] | [0.012] | [0.01] | | | [---] | [0.071] | [3.23] |
| **DF** | _ | 199.99*** | _ | _ | α | 0.0000 | 0.000 | -0.007 | -1.84 |
| | | [2.69] | | | | [---] | [11.28] | [0.017] | [2.83] |
| **AIC** | -8.94 | -6.76 | -0.812 | 1.81 | β | 0.23 | 0.67 | 0.55*** | 0.171 |
| | | | | | | [---] | [179.66] | [0.21] | [0.72] |
| **LL** | 4.4733 | 4.38 | 9.655 | 0.091 | AIC | -4.94 | -2.79 | **-21.4** | 5.615 |
| | | | | | LL | 4.47 | 4.4 | -10.743 | 0.192 |

Note: This Table reports the obtained results of static and dynamic copula model estimates for the dependence between JPY/USD exchange rate and the global EU. CST is the constant parameter of the copula. DF is the degree of freedom of student-t copula. The value in bold reflects the lower AIC. ** and *** indicate statistical significances at 5% and 1% levels respectively. Standard errors are given in brackets. [---] denotes large number.

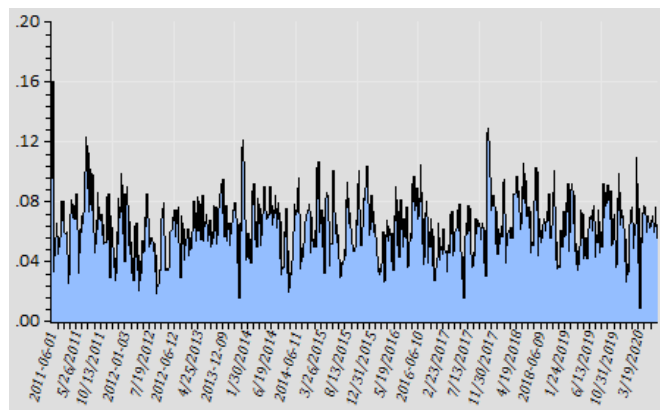

Figure B.2. Tail Dependence of the Japanese currency return with the global EU (Gumbel Copula)